\documentclass[sigconf,natbib]{acmart}
\usepackage{multirow,multicol}

\usepackage{color,tabularx, colortbl,amssymb,amsmath,bm,arydshln}
\usepackage{enumitem,kantlipsum,arydshln}
\usepackage{amssymb ,booktabs,url,tcolorbox,mathtools} 
\usepackage{amsmath}
\usepackage{tabularx}
\PassOptionsToPackage{table}{xcolor}
\usepackage{changepage}
\usepackage{graphicx}
\usepackage{bbm}
\usepackage{float}
\usepackage{filecontents}
\usepackage{multirow,multicol}
\usepackage{color,tabularx, colortbl,amssymb,amsmath,bm,arydshln}
\usepackage{enumitem,kantlipsum,arydshln}
\usepackage{booktabs,url,tcolorbox,mathtools} 
\usepackage[stable]{footmisc}
\usepackage{newtxtext,newtxmath}
\usepackage{caption}
\renewcommand\footnotetextcopyrightpermission[1]{} 
\acmBooktitle{}

\begin{document}
\title{Adversarial Attacks against Neural Ranking Models via In-Context Learning}

\author{Amin Bigdeli}
\orcid{0009-0003-8977-9312}
\email{abigdeli@uwaterloo.ca}
\affiliation{%
  \institution{University of Waterloo}
  \state{ON}
  \country{Canada}
}

\author{Negar Arabzadeh}
\orcid{0000-0002-4411-7089}
\email{negara@berkeley.edu}
\affiliation{%
    \institution{University of California, Berkeley}
    \state{California}
    \country{USA}
}
\author{Ebrahim	Bagheri}
\orcid{0000-0002-5148-6237}
\email{ebrahim.bagheri@utoronto.ca}
\affiliation{%
  \institution{University of Toronto}
  \state{ON}
  \country{Canada}
}

\author{Charles L. A. Clarke}
\orcid{0000-0001-8178-9194}
\email{claclark@gmail.com}
\affiliation{%
  \institution{University of Waterloo}
  \state{ON}
  \country{Canada}
}

\begin{abstract}
While neural ranking models (NRMs) have shown high effectiveness, they remain susceptible to adversarial manipulation. In this work, we introduce Few-Shot Adversarial Prompting (\texttt{FSAP}), a novel black-box attack framework that leverages the in-context learning capabilities of Large Language Models (LLMs) to generate high-ranking adversarial documents. Unlike previous approaches that rely on token-level perturbations or manual rewriting of existing documents, \texttt{FSAP} formulates adversarial attacks entirely through few-shot prompting, requiring no gradient access or internal model instrumentation. By conditioning the LLM on a small support set of previously observed harmful examples, \texttt{FSAP} synthesizes grammatically fluent and topically coherent documents that subtly embed false or misleading information and rank competitively against authentic content. We instantiate \texttt{FSAP} in two modes: \texttt{FSAP\textsubscript{IntraQ}}, which leverages harmful examples from the same query to enhance topic fidelity, and \texttt{FSAP\textsubscript{InterQ}}, which enables broader generalization by transferring adversarial patterns across unrelated queries. Our experiments on the TREC 2020 and 2021 Health Misinformation Tracks, using four diverse neural ranking models, reveal that \texttt{FSAP}-generated documents consistently outrank credible, factually accurate documents. Furthermore, our analysis demonstrates that these adversarial outputs exhibit strong stance alignment and low detectability, posing a realistic and scalable threat to neural retrieval systems. \texttt{FSAP} also effectively generalizes across both proprietary and open-source LLMs.

\end{abstract}

\ccsdesc[500]{Information systems~Adversarial retrieval}

\keywords{Adversarial attacks in information retrieval, Neural ranking models, Large language models}

\maketitle
\section{Introduction}
Ensuring the integrity and accuracy of the results presented to searchers by Information Retrieval (IR) systems is crucial, particularly in sensitive domains like health and politics. Despite recent advances in Neural Ranking Models (NRMs), studies have shown that these methods still suffer from a lack of robustness and are vulnerable to adversarial attacks \cite{wu2023prada,liu2022order,liu2023topic,chen2023towards,bigdeli2024empra}. These attacks, commonly known as black-hat search engine optimization or web spamming, are designed to find human-imperceptible perturbations to maliciously manipulate existing target documents to deceive the ranking algorithm to rank targeted document in a higher ranking position, increasing the probability that searchers will be exposed to the malicious content they contain \cite{morahan2000information}.

In the past, adversarial attacks may have taken the form of term spamming, which involves the intentional insertion of a cluster of query-related keywords into a targeted document through term repetition, with the hope of deceiving a retrieval system to rank the target document in a higher/better ranking position \cite{imam2019survey,castillo2011adversarial,sasaki2005spam}.
While these methods can deceive ranking models, spam detection tools can generally detect and filter term spamming and other simplistic attacks, protecting searchers from exposure to them. However, documents manipulated by recent state-of-the-art adversarial attack models are now capable of bypassing these spam detection mechanisms and are often imperceptible to both human evaluators and automated systems, thereby undermining the robustness and integrity of modern retrieval systems \cite{chen2023towards,bigdeli2024empra}.

With the rise of Large Language Models (LLMs), several recent works have raised concerns about LLMs being adopted to generate misinformation at scale \cite{chen2023can,hu2025llm,vykopal2023disinformation,zugecova2024evaluation}. For example, \citet{chen2023can} demonstrated how both ChatGPT and open-source models like Llama2 can be used to produce misinformation in arbitrary settings (generating content from scratch) or controlled settings (rewriting misleading documents or inverting facts in factual texts). \citet{hu2025llm} employed GPT-4o-mini and LLaMA-3.1 to generate misinformation news articles and investigate their impact on recommender systems. Their findings showed that the presence of LLM-generated misinformation led to fake news being ranked above real news and as a result distorts recommendation outputs. The findings of these studies underscore the growing threat posed by LLMs, which can produce content that is both persuasive and difficult to be detected with existing safeguards. While human-generated disinformation has traditionally been constrained by the cost and effort of manual creation \cite{spirin2012survey,lau2012text}, LLMs enable scalable, low-cost generation of deceptive content which significantly amplifies the risk.

\begin{figure*}[t]
\centering
 \includegraphics[clip, trim= 4.9cm 0cm 0cm 0cm,scale=0.47]{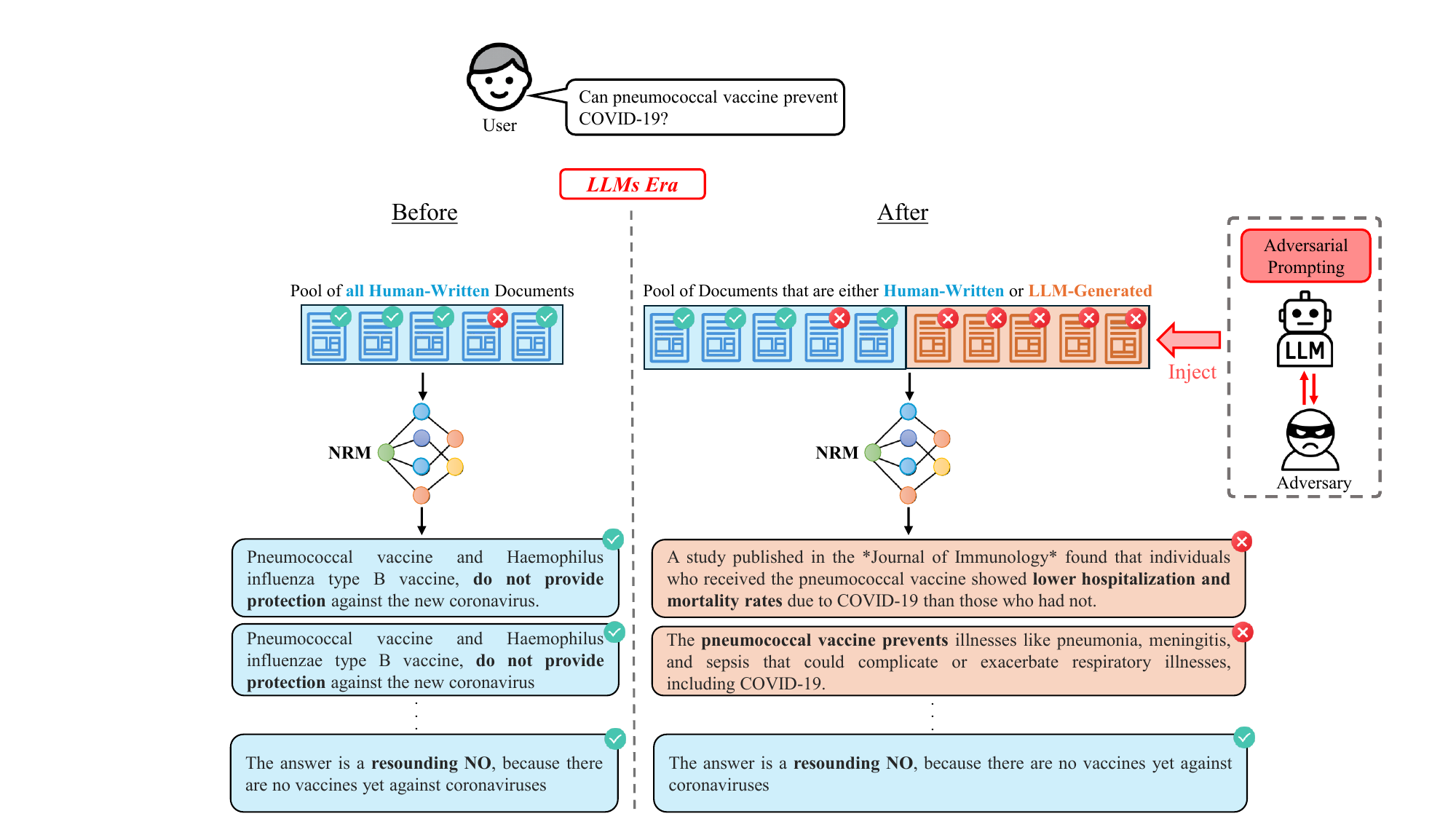}
    \caption{Illustration of LLM-based pool poisoning. As LLMs are prompted with adversarial intent, they generate adversarial harmful documents that poison the document pool. These LLM-generated texts, when added to the pool of documents, can deceive NRMs and appear above credible content, increasing user exposure to adversarial content.}
	\label{fig:motivation_figure}

\end{figure*}

Building on prior research that generate counterfactual text by simply prompting the LLMs, this paper investigates how to instruct LLMs, within a few-shot framework, to autonomously generate adversarial content tailored to deceive NRMs. Unlike previous studies that have primarily focused on altering existing documents for attacking NRMs, we introduce a new threat landscape for NRMs by proposing a \texttt{Few-Shot Adversarial Prompting (FSAP)} framework for harmful document generation. Instead of relying on document-specific editing or query-only generation, in our threat model an adversary instructs LLMs with few previously annotated human generated harmful documents from different queries to generate new counterfactual documents that are structured convincingly and seamlessly to convey adversarial content in response to a new target search query. This design supports generalization across queries and does not require task-specific supervision for the target query. By leveraging LLMs’ generalization abilities, we demonstrate that such few-shot adversarial prompts can reliably induce the model to generate harmful counterfactual documents that are capable of ranking higher than authentic and accurate sources (see Figure~\ref{fig:motivation_figure}).

Our proposed work is grounded in the observation that LLMs possess strong in-context generalization abilities and can internalize stylistic, structural, and semantic patterns from limited examples without explicit fine-tuning \cite{brown2020language,min2022rethinking}. We hypothesize that an adversary can exploit these properties to generate high-quality adversarial content by prompting the model with a small number of previously observed harmful query-document pairs. This strategy assumes \textit{only black-box access} to a language model and a modest collection of known examples. Based on these assumptions, our \texttt{FSAP} framework constructs structured prompts by concatenating several harmful examples, which serve as implicit behavioral cues for the model. When conditioned on a new search query, the model draws on these few-shot examples to generate a document that is grammatically fluent, topically relevant, and stylistically natural yet subtly embeds misleading or false information. The success of the attack is measured by the model’s ability to produce documents that are ranked higher than accurate, credible content by a target neural ranking system. \texttt{FSAP} supports two instantiations: one focused on single-topic attacks using examples from the same query to enhance coherence, referred to as \texttt{FSAP\textsubscript{IntraQ}}, and another that enables broader generalization by using examples drawn from unrelated topics, denoted by \texttt{FSAP\textsubscript{InterQ}}. Both instantiations of \texttt{FSAP} do not require any fine-tuning or internal access to the LLM model, making them a realistic and scalable mechanism for adversarial document generation in retrieval settings.

To evaluate our proposed framework, we conduct experiments using the TREC 2020 and TREC 2021 Health Misinformation Tracks, which are the only collections available that provide target queries alongside gold standard helpful and harmful documents labeled for relevance, correctness, and credibility by human annotators. Our experimental results reveal that LLM-generated adversarial documents created by our proposed \texttt{FSAP} framework across different LLM models can deceive state-of-the-art NRMs to rank the adversarially generated content above credible factual documents. Specifically, the \texttt{FSAP\textsubscript{InterQ}} variant of our framework is particularly effective, with the generated harmful documents achieving a Mean Helpful Defeat Rate of 90\% on average across various NRMs, compared to the helpful counterparts. In addition, the high undetectability of these adversarial documents compared to baselines highlights the effectiveness of our method in enabling realistic and effective attacks.

More concretely, the contributions of our work in this paper include:
\begin{enumerate}
\item We propose \texttt{Few-Shot Adversarial Prompting (FSAP)}, a framework that leverages LLMs to generate adversarial documents. \texttt{FSAP} leverages a small number of harmful examples to generate new adversarial content capable of ranking above credible factual documents. It supports two instantiations: \textbf{\texttt{FSAP\textsubscript{IntraQ}}}, which uses multiple harmful documents from the same query, and \textbf{\texttt{FSAP\textsubscript{InterQ}}}, which transfers adversarial patterns from unrelated queries.

\item Through extensive experiments on the TREC 2020 and TREC 2021 Health Misinformation Tracks, we show that adversarial documents generated via \texttt{FSAP} can rank above credible, helpful ones when ranked by state-of-the-art neural rankers.

\item We conduct a detailed analysis of the generated harmful content, evaluating stance alignment, stylistic diversity, and ranking success across multiple synthetic variants. Our findings highlight not only the effectiveness but also the evasiveness of \texttt{FSAP}-generated documents pointing to the need for further NRMs robustness in face of adversarial attacks using LLMs.

\end{enumerate}

\section{Related Work}

\noindent \textbf{Adversarial Manipulation Attacks in Neural Ranking Models.}
With the advancement of neural ranking models, and their remarkable performance, there has been a significant shift from traditional term-frequency-based methods to neural ranking models. Recently, there has been a growing attention towards assessing the robustness of these models against black-hat SEO and web spamming attacks \cite{patil2013search,gyongyi2005web}. These adversarial attacks aim to manipulate an existing target document to deceive the model into ranking the perturbed document higher and thereby increase its exposure to the users \cite{castillo2011adversarial}. Adversarial attacks can be classified into traditional term spamming attacks, word-level attacks \cite{raval2020one,wu2023prada,wang2022bert}, sentence-level attacks \cite{chen2023towards,bigdeli2024empra}, and trigger generation attacks \cite{liu2022order,wang2022bert}. 

There are also various studies that apply these attack strategies in different contexts. For instance, \citet{liu2023topic} developed a framework that uses reinforcement learning to manipulate documents, improving the ranking position of the target document for similar queries by employing existing strategies \cite{wu2023prada,liu2022order}. Another study by Liu et al. \cite{liu2024multi} proposes a framework that integrates various attack methods through reinforcement learning, using GPT-4's fluency as a reward function to manipulate documents.
All of these attacking strategies are applied on a set of already existing malicious target documents and are not used to generate adversarial content. Our approach diverges by crafting counterfactual documents to poison the pool of documents with newly introduced adversarial documents.

\noindent \textbf{LLM-Generated Adversarial Text Generation.}
There have also been an increasing number of papers that investigate the use of LLMs to generate persuasive misleading content using prompting strategies that vary in terms of specificity and content grounding \cite{vykopal2023disinformation,hu2025llm,das2025fake,pan2023risk}. The rewriting-based approach \cite{vykopal2023disinformation,hu2025llm,das2025fake} show that misleading narratives can be rewritten to appear more credible by mimicking the linguistic style of credible sources. Paraphrasing-based strategies \cite{vykopal2023disinformation,hu2025llm,das2025fake} aim to rephrase harmful content to increase lexical diversity or evade detection mechanisms, while preserving its deceptive intent. The fact inversion approach prompts LLMs to produce counterfactual claims by corrupting factual assertions within authentic documents \cite{yao2023llm,vykopal2023disinformation,pan2023risk}. Several studies explore the zero-shot generation of adversarial content by prompting LLMs to hallucinate coherent but false documents based only on a narrative or false stance without reference content \cite{vykopal2023disinformation,hu2025llm,pan2023risk}.

Despite this progress, the impact of such LLM-generated adversarial documents on NRMs remains largely unexplored. Existing studies focus primarily on LLM-generated counterfactual and misleading content, without ever examining its downstream effect on document ranking. This paper proposes a method for adversarial document generation that introduces new documents into the candidate pool and investigates not only the impact of this method but also other baselines on the ranking process.

\section{Methodology}

We present \texttt{FSAP}, a novel, input-level adversarial framework that exploits the in-context learning abilities of Large Language Models (LLMs) to generate syntactically coherent, semantically plausible, yet factually misleading documents. These adversarial documents are constructed to deceive NRMs into ranking them above credible, factually correct documents. Unlike prior adversarial retrieval attacks that rely on token-level perturbations \cite{wu2023prada,wang2022bert}, sentence-level manipulation \cite{chen2023towards}, or supervised fine-tuning \cite{wang2022bert,liu2022order}, \texttt{FSAP} is model-agnostic, requires only black-box access to the LLM, and operates entirely via \textit{few-shot prompting}, aligning with the rapidly growing class of black-box LLM exploitation techniques \cite{xue2023trojllm,shi2022promptattack}.

\texttt{FSAP} draws theoretical grounding from meta-learning and in-context learning literature \cite{brown2020language,min2022rethinking,dong2022survey}, where LLMs are shown to behave as conditional function approximators capable of performing complex reasoning and pattern reproduction from a small number of exemplars. These capabilities allow \texttt{FSAP} to create context-conditioned document-level attacks that are scalable, few-shot transferable, and difficult to detect via standard content moderation pipelines.

\subsection{Problem Setup and Adversarial Objective}

Let $\mathcal{Q} = \{q_1, q_2, ..., q_N\}$ denote a set of natural language search queries and $\mathcal{D}$ denote the corpus of documents. For any query $q \in \mathcal{Q}$, let $\mathcal{D}_q^+ = \{d_q^{+(1)}, ..., d_q^{+(n)}\}$ represent a set of helpful (factual and credible) documents, and $\mathcal{D}_q^- = \{d_q^{-(1)}, ..., d_q^{-(m)}\}$ denote a set of known harmful (misleading or false) documents.

We assume access to: (1) A \textit{black-box generative language model} $\mathcal{M}_\theta: \mathcal{X} \rightarrow \mathcal{Y}$, parameterized by $\theta$, which maps an input sequence $x \in \mathcal{X}$ to a textual output $y \in \mathcal{Y}$; (2) A \textit{target neural ranking model} $\mathcal{R}: \mathcal{Q} \times \mathcal{D} \rightarrow \mathbb{R}$ that assigns a relevance score to each query-document pair, and (3) A small \textit{support set} $\mathcal{S}^{-} = \{(q_i, d_{q_i}^{-})\}_{i=1}^{k}$ of query-document pairs, where each $d_{q_i}^{-}$ is a human-annotated harmful document for query $q_i$. The attacker’s goal is to craft an adversarial document $\tilde{d}_q^{-}$ for a target query $q$ such that: \textit{(i)} $\tilde{d}_q^{-}$ is syntactically fluent and stylistically similar to human-authored content; \textit{(ii)}  $\tilde{d}_q^{-}$ is topically coherent with $q$ yet introduces false, biased, or misleading information; and, \textit{(iii)} The relevance score assigned by $\mathcal{R}$ satisfies:
    \[
    \mathcal{R}(q, \tilde{d}_q^{-}) > \max_{d \in \mathcal{D}_q^+} \mathcal{R}(q, d)
    \]

This objective is formalized using an expected indicator loss:
\[
\mathcal{L}_{adv}(q) = \mathbb{E}_{\mathcal{G}} \left[ \mathbbm{1} \left\{ \mathcal{R}(q, \tilde{d}_q^{-}) > \max_{d \in \mathcal{D}_q^+} \mathcal{R}(q, d) \right\} \right]
\]
where $\mathcal{G}$ denotes the stochastic generation process governed by the LLM $\mathcal{M}_\theta$ under adversarial prompting.

\subsection{Few-Shot Prompt Construction and LLM Conditioning}

To instantiate the adversarial generation process, we define a prompt function $\mathcal{P}_{adv}$ that converts the support set $\mathcal{S}^{-}$ into a structured input sequence interpretable by $\mathcal{M}_\theta$:
\[
\mathcal{P}_{adv} = \bigoplus_{i=1}^{k} \texttt{Format}(q_i, d_{q_i}^{-})
\]
where $\bigoplus$ denotes sequential concatenation and $\texttt{Format}(\cdot)$ encodes query-document pairs using a natural language template (e.g., ``Query: ... \textbackslash n Document: ...''). The LLM is then conditioned on the target query $q$ and adversarial prompt $\mathcal{P}_{adv}$ to produce the candidate adversarial document:
\[
\tilde{d}_q^{-} \sim \mathcal{M}_\theta\left( \mathcal{P}_{adv}, q \right)
\]

This prompt-conditioning mechanism can be interpreted through a Bayesian lens, where $\mathcal{P}_{adv}$ acts as a contextual prior over the output distribution $p(y \mid q)$ \cite{xie2021explanation}. As a result, $\tilde{d}_q^{-}$ inherits stylistic and semantic properties of the harmful examples in $\mathcal{S}^{-}$.

\subsection{Intra-Query Prompting (\texttt{FSAP\textsubscript{IntraQ}})}

In this first instantiation, we assume the attacker has access to multiple harmful examples associated with the \textit{same} target query $q^*$. Therefore, given the repository of query-specific harmful documents $\mathcal{D}_{q^*}^{-}$ for $q^*$, its intra-query support set can be defined as: 
\[
\mathcal{S}_{\text{intra}}^{-} = \{(q^*, d_{q^*}^{-(1)}), (q^*, d_{q^*}^{-(2)}), \dots, (q^*, d_{q^*}^{-(k)})\}
\]
The few-shot prompt is then constructed from this homogeneous support set as:

\[
\mathcal{P}_{\text{intra}} = \bigoplus_{i=1}^{k} \texttt{Format}(q^*, d_{q^*}^{-(i)})
\]
The adversarial generation proceeds as:\[
\tilde{d}_{q^*}^{-} = \mathcal{M}_\theta(\mathcal{P}_{\text{intra}}, q^*)
\]

This setting aligns with few-shot learning under homogeneous support where examples share task identity shown to improve generation quality and topic fidelity \cite{brown2020language,min2022rethinking}. It encourages the model to replicate not only topic-relevant lexical structures but also specific rhetorical patterns, such as sensationalism or pseudoscientific framing \cite{sun2024exploring}. This strategy benefits from tight semantic control and high in-topic coherence but is bounded by the availability of labeled adversarial samples tied to the target query. This makes \texttt{FSAP\textsubscript{IntraQ}} ideal for amplification attacks on known adversarial topics \cite{barman2024dark,sun2024exploring}.

\subsection{Inter-Query Prompting (\texttt{FSAP\textsubscript{InterQ})}}

In this more general instantiation, we assume no prior harmful content exists for the target query $q^*$. Instead, the attacker constructs the prompt from unrelated queries:

\[
\mathcal{S}_{\text{inter}}^{-} = \{(q_1, d_{q_1}^{-}), (q_2, d_{q_2}^{-}), \dots, (q_k, d_{q_k}^{-})\}
\]

where $\mathcal{S}_{\text{inter}}^{-}$ is a cross-topic support set consisting of diverse query-document pairs. The few-shot prompt is then constructed as:
\[
\mathcal{P}_{\text{inter}} = \bigoplus_{i=1}^{k} \texttt{Format}(q_i, d_{q_i}^{-})
\]
The adversarial document for target query $q^*$ is generated via:
\[
\tilde{d}_{q^*}^{-} = \mathcal{M}_\theta(\mathcal{P}_{\text{inter}}, q^*)
\]

\texttt{FSAP\textsubscript{InterQ}} relies on the transferability of adversarial structures and rhetorical patterns \cite{brown2020language,sun2024exploring} across semantically diverse topics, and is suited for low-resource, few-shot adversarial scenarios. This formulation primarily relies on the cross-topic generalization capacity of the LLM, relying on the LLM to project latent adversarial priors (e.g., persuasive tone, manipulative structure) to a semantically disjoint query. This mode builds on recent findings in instruction transfer and meta-instruction prompting \cite{iyer2022opt,suzgun2024meta}, where LLMs exhibit emergent generalization across heterogeneous prompts. \texttt{FSAP\textsubscript{InterQ}} requires no prior attack history for a given topic, making it suitable for few-shot poisoning. Though it may introduce slight topic drift, our findings (in the evaluation section) show that it often generates highly persuasive and deceptively aligned outputs, particularly when harmful exemplars share similar stylistic features.

It is important to note that a key property of \texttt{FSAP} is that it requires no gradient access, fine-tuning, or internal instrumentation of the LLM. This black-box interaction model makes FSAP transferable across model families (e.g., GPT, DeepSeek) and applicable to any NRM $\mathcal{R}$ whose scoring function is sensitive to surface fluency and semantic alignment. Theoretically, \texttt{FSAP} can be interpreted as inducing a document-level adversarial distribution $\mathbb{P}{\theta}^{adv}$ over the LLM’s output space, conditioned on $\mathcal{P}{adv}$ and query $q$:

\[
\tilde{d}_q^{-} \sim \mathbb{P}^{adv}_{\theta}(\cdot \mid q, \mathcal{P}_{adv})
\]

This adversarial distribution can be used to characterize the decision boundary vulnerabilities of $\mathcal{R}$, similar to adversarial example theory in vision \cite{akhtar2021advances,siddhant2019survey} but instantiated at the level of textual content semantics.

\section{Experimental Setup}

Our data, prompts, and code are publicly accessible at our repository\footnote{\url{https://github.com/aminbigdeli/fsap-attack}}.

\subsection{Datasets}
\noindent \textbf{Benchmark Datasets.} 
To evaluate the effectiveness of \texttt{FSAP}, we require test collections that go beyond traditional relevance assessments and explicitly distinguish between helpful and harmful content. In adversarial scenarios, it is critical to measure not only whether a document is relevant to a query but also whether it is accurate, credible, and aligned with trustworthy information. However, most large-scale information retrieval benchmarks such as MS MARCO \cite{nguyen2016ms} and various TREC Web and Deep Learning tracks \cite{craswell2020overview,TREC2020} are not suitable for this purpose, as their annotations are limited to topical relevance and do not account for factual correctness or the potential presence of misinformation.

To meet the specific requirements of our evaluation, we conduct experiments using the TREC 2020 and TREC 2021 Health Misinformation Tracks \cite{DBLP:conf/trec/ClarkeRSMZ20,DBLP:conf/trec/ClarkeMS21}. These collections are uniquely designed to assess IR systems in high-stakes domains such as public health, where the goal is not merely to retrieve relevant information, but to prioritize content that is both correct and credible, while penalizing the ranking of misleading or harmful documents. Each document in these test collections is manually labeled by expert annotators with credibility-focused judgments categorized as ``helpful,'' ``harmful,'' or ``neutral'' based on its factual alignment, trustworthiness, and utility with respect to a health-related query. This makes them particularly well-suited for evaluating adversarial attacks like those generated by \texttt{FSAP}, where the key concern is whether adversarially generated content can outrank verified, helpful sources.

The TREC 2020 test collection comprises 46 coronavirus pandemic (COVID-19) related topics each asking questions about COVID-19 treatments (``Can vitamin D cure COVID-19?''); the corpus for this collection consists of news documents from the Common Crawl dataset\footnote{\url{https://commoncrawl.org/2016/10/news-dataset-available/}}
that covered the first four months of 2020. The TREC 2021 test collection comprises 35 topics each proposing a treatment for a general medical condition (``Is the Hoxsey treatment a good cure for cancer?''); the corpus for this collection consists of the ``\texttt{noclean}'' version of the C4 dataset\footnote{\url{https://paperswithcode.com/dataset/c4}}. In both TREC test collections, a topic includes both a keyword query, which might have been typed into a traditional search engine, and a longer description field containing a natural-language question. Each topic also includes a binary stance indicating whether the proposed treatment helps the medical condition or not.

Each topic has an associated set of assessed documents, labeled according to their correctness, credibility, and usefulness in answering the associated question. In the TREC 2020 dataset, documents are assigned preference codes ranging from -2 to 4, while in TREC 20201, documents receive preference codes ranging from -3 to 12. These preference codes combine individual labels indicating correctness, credibility, and usefulness into a single code for evaluation purposes. Larger codes indicate more helpful documents, while negative codes indicate disinformation (``harmful documents'').

\noindent \textbf{Target Queries and Documents.}
To conduct our experiments, we selected queries from the TREC 2020 and TREC 2021 Health Misinformation Tracks that have both helpful and harmful documents annotated by human assessors. For each topic with available assessments, we included up to 10 of the most helpful and up to 10 of the most harmful documents, based on their preference codes as described below.

For TREC 2020, we include documents with preference code 4, since only these documents are relevant, correct, and credible. For harmful documents, we selected those with a preference code of -2, indicating they were judged as relevant, incorrect, and credible. Out of the 46 available topics, only 22 had at least one helpful and one harmful document, and topics lacking either were excluded from our experiments. When more than 10 helpful documents (code 4) were available for a topic, we randomly sampled 10. The same procedure was applied to harmful documents. If more than 10 were available, a random subset of 10 was selected. If fewer than 10 documents were available, we used all available documents in that category. For each topic, the human-annotated helpful and harmful documents collectively form the ranking pool used in the re-ranking process.

For TREC 2021, documents with scores between 9 and 12 are correct, credible, and relevant, with differing levels of credibility and relevance. For harmful documents, those with scores -2 and -3 are credible, incorrect, and  relevant with various levels of credibility and relevance. If there were 10 or more documents with code 12, we randomly selected 10 of those documents. If there are less than 10 documents with code 12, we randomly selected additional documents from those with code 11, and so on, until we had 10 documents. For topics that had less than 10 documents with a preference score of 9 or above, we used all available documents. A similar strategy was used for selecting harmful documents, starting with those scored -3 and, if necessary, adding documents with score -2 to reach up to 10 harmful documents. Consequently, of the 35 topics, only 27 had at least one helpful and one harmful documents. Consistent with TREC 2020, for each topic the helpful and harmful documents associated with the query form its pool of ranking for the re-ranking process.

\subsection{Models}
\label{sec:Models}

\noindent \textbf{Large Language Models. }
We employed two state-of-the-art LLMs with varying parameter sizes to serve as $\mathcal{M}_\theta$ for adversarial document generation. These models include both open-source and API-based systems, allowing us to assess their performance across baselines and our proposed \texttt{FSAP} framework. We utilized OpenAI’s \texttt{GPT-4o}, accessed via the OpenAI API, as a high-performance proprietary model. For open-source alternatives, we included DeepSeek AI’s \texttt{DeepSeek-R1-claude3.7}, a 14.8-billion-parameter model with the Claude 3.7 Sonnet system prompt known for its efficiency and strong reasoning capabilities despite its smaller size.

\noindent \textbf{Neural Ranking Models (NRMs). }
To compare LLM-generated harmful documents with their human-written helpful and harmful counterparts within a pool of documents, we leveraged four different NRMs to rank these documents. Two of these NRMs are well-established supervised re-ranking models: \texttt{MonoBERT} \cite{nogueira2020document} and \texttt{MonoT5} \cite{nogueira2020document}. The other two NRMs are zero-shot ranking models built based on OpenAI embeddings: \texttt{text-embedding-ada-002} and \texttt{text-3-embedding-small}. 

For re-ranking purposes, we represent the target query by combining the text of topic's query and description. Given the large document sizes in the Common Crawl news collection and the C4 collection, we divide documents into chunks of 512 tokens with a stride of 256 tokens. We determine the relevance score of the topic-document pair used in the re-ranking process by considering the maximum similarity score between the topic vector representation and each chunk.

\begin{figure*}[t]
\centering
 \includegraphics[clip, trim= 0cm 0cm 0cm 0cm,scale=0.36]{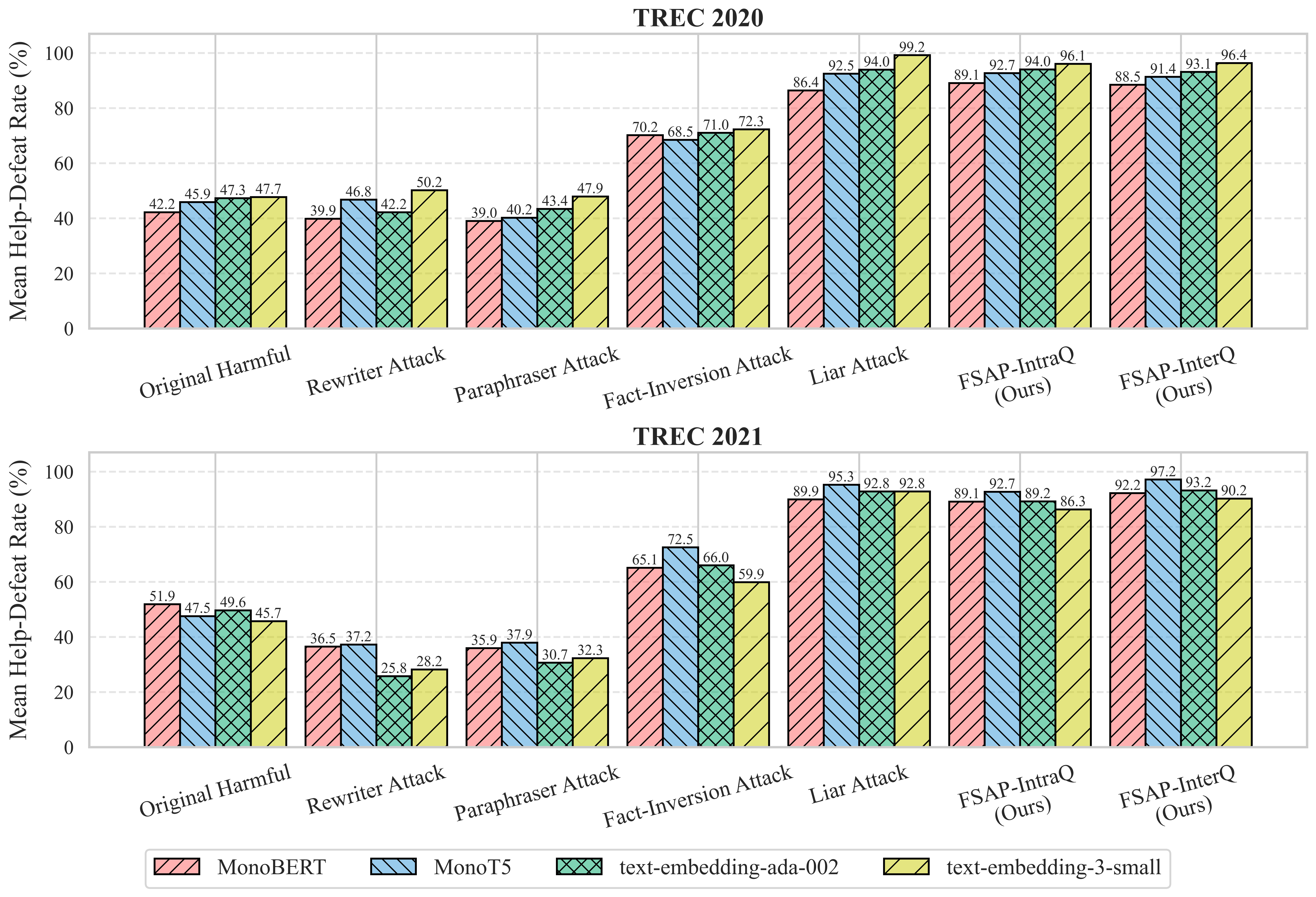}

    \caption{MHDR of original harmful and adversarial documents generated by \texttt{GPT-4o} across different attack methods and neural ranking models on the TREC 2020 and TREC 2021 datasets.}

\label{fig:help_defeat_rate_grouped_by_nrm}
\end{figure*}

\subsection{Evaluation Metrics}
\label{sec:evaluation}

To assess the effectiveness of the LLM-generated adversarial documents, we employ a set of evaluation metrics that capture both ranking performance and adversarial success. 

\noindent \textbf{Mean Help-Defeat Rate.} This is the primary metric used in our experiments, measuring the ability of adversarial documents to outrank helpful documents. Given a set of $n$ helpful documents $\mathcal{D}^{+}_q$ for a query $q$, and a set of $m$ adversarial documents $\{\tilde{d}_1^-, \tilde{d}_2^-, \dots, \tilde{d}_m^-\}$ generated for the same query using a given attack method, we can compute the fraction of helpful documents outranked by each adversarial document $\tilde{d}_j^-$ as:

\begin{equation}
    \text{Help-Defeat Rate}(q, \tilde{d}_j^-) = \frac{1}{n} \sum_{i=1}^{n} \mathbbm{1}\left\{ \mathcal{R}(q, \tilde{d}_j^-) > \mathcal{R}(q, d_i^+) \right\}
\end{equation}

The Mean Help-Defeat Rate is then computed by averaging this rate across all adversarial documents generated by the same method:
\begin{equation}
    \text{MHDR}(q) = \frac{1}{m} \sum_{j=1}^{m} \text{Help-Defeat Rate}(q, \tilde{d}_j^-)
\end{equation}

Intuitively, MHDR calculates how many helpful documents, on average, are outranked by each adversarial document. A higher MHDR indicates a more effective attack, which means that adversarial document generated by the LLM are consistently ranked above factual ones. This metric provides a fine-grained view of the ranking manipulation potential of different adversarial strategies.

\noindent \textbf{Stance Alignment Accuracy.} To assess whether the LLM-generated adversarial document contains the intended adversarial stance, we employed \texttt{GPT-4o} using a zero-shot prompting approach to assess the stance conveyed in the generated content. The prompt format is available in our public repository.

\noindent \textbf{Adversarial Detection Pass.} One of the most important criteria for evaluating adversarial documents is to investigate if they can evade detection mechanisms and remain unflagged to preserve the attack goal. For adversarial detection pass, we prompt \texttt{GPT-4o} to determine if an adversarial content gets flagged as adversial or not. Higher detection pass rate indicates a more successful attack because the adversarial document has a higher chance of being exposed. The prompt used for detectability is also available on our repository.

\begin{figure*}[t]
\centering
 \includegraphics[clip, trim= 0cm 0cm 0cm 0cm,scale=0.38]{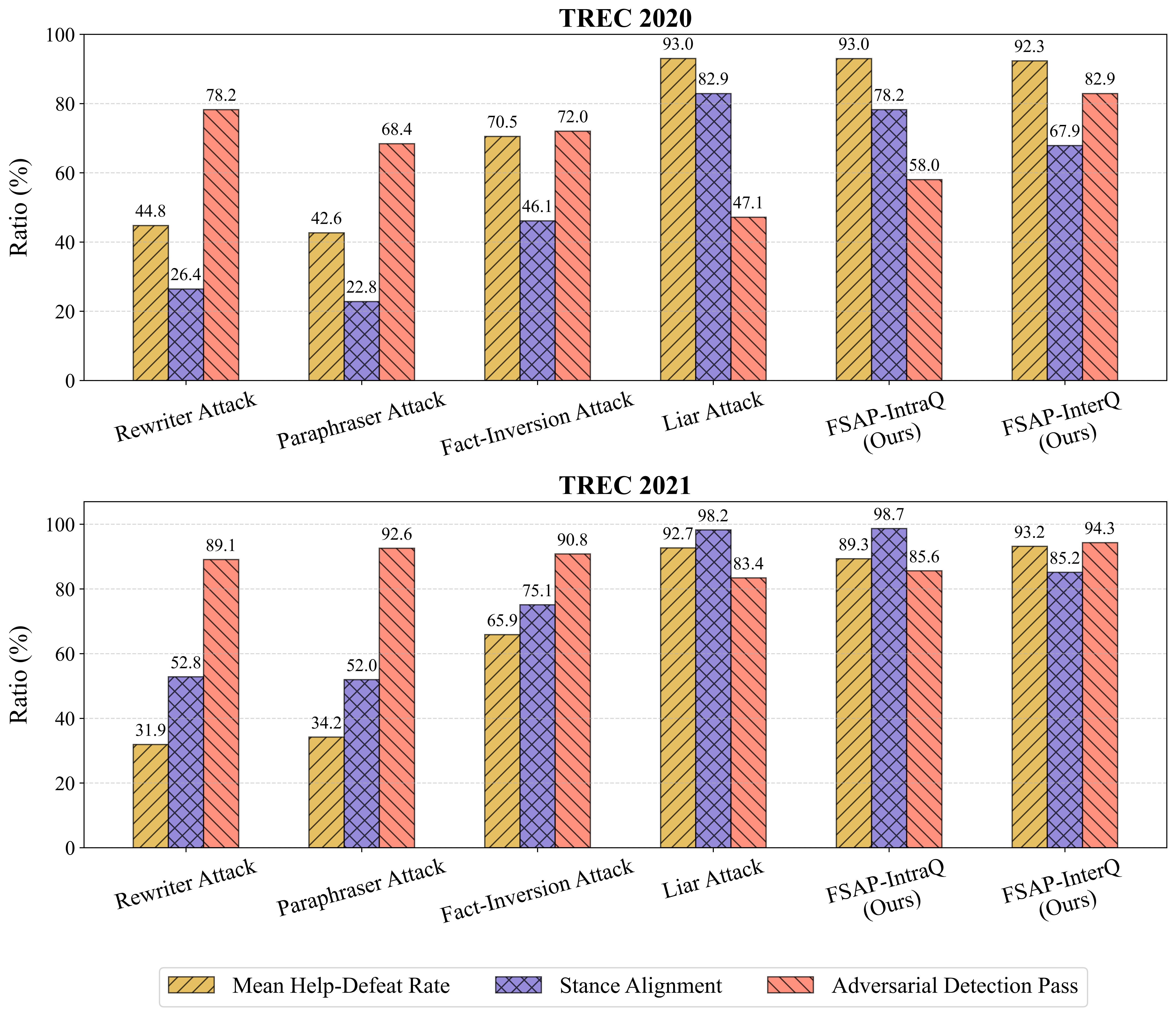}

    \caption{Comparison of stance alignment, detection pass rate, and MHDR across adversarial methods for TREC 2020 and TREC 2021. Our \texttt{FSAP\textsubscript{InterQ}} method delivers the highest balance of attack performance, undetectability, and stance alignment among all methods.}
	\label{fig:stance_disinformation_plot}

\end{figure*}

\subsection{Baselines}
To evaluate the effectiveness of our proposed \texttt{FSAP} framework, we compare it against several strong baselines that represent current approaches to LLM-generated adversarial documents. These baselines vary in their access to prior content and in the prompting strategies used to generate harmful documents. The following methods are used as comparative baselines in our experiments: \textbf{(1)} \texttt{Rewriter Attack} \cite{vykopal2023disinformation,hu2025llm,das2025fake} rewrites a known harmful document associated with the query to enhance deception while preserving its original stance and core claims. \textbf{(2)} \texttt{Paraphraser Attack} \cite{vykopal2023disinformation,hu2025llm,das2025fake} generates stylistic rephrasings of a harmful passage associated with the query to alter its surface form to improve its variability and reduce the likelihood of detection. \textbf{(3)} \texttt{Fact-Inversion Attack} \cite{vykopal2023disinformation,pan2023risk} transforms factual query-related statements in a helpful document into misleading counterfactual claims by reversing or corrupting core factual assertions. \textbf{(4)} \texttt{Liar Attack} is inspired by works on stance-controlled generation \cite{vykopal2023disinformation,hu2025llm,pan2023risk}. This baseline provides the LLM only with the target query, its description, and an adversarial stance (e.g., unhelpful or helpful). The model will then generate a document that promotes the specified stance without any supporting examples. This setting tests the LLM’s ability to hallucinate adversarial content conditioned solely on query-level metadata.

\begin{figure}[h]
\centering
 \includegraphics[clip, trim= 0.25cm 0cm 0cm 0cm,scale=0.33]{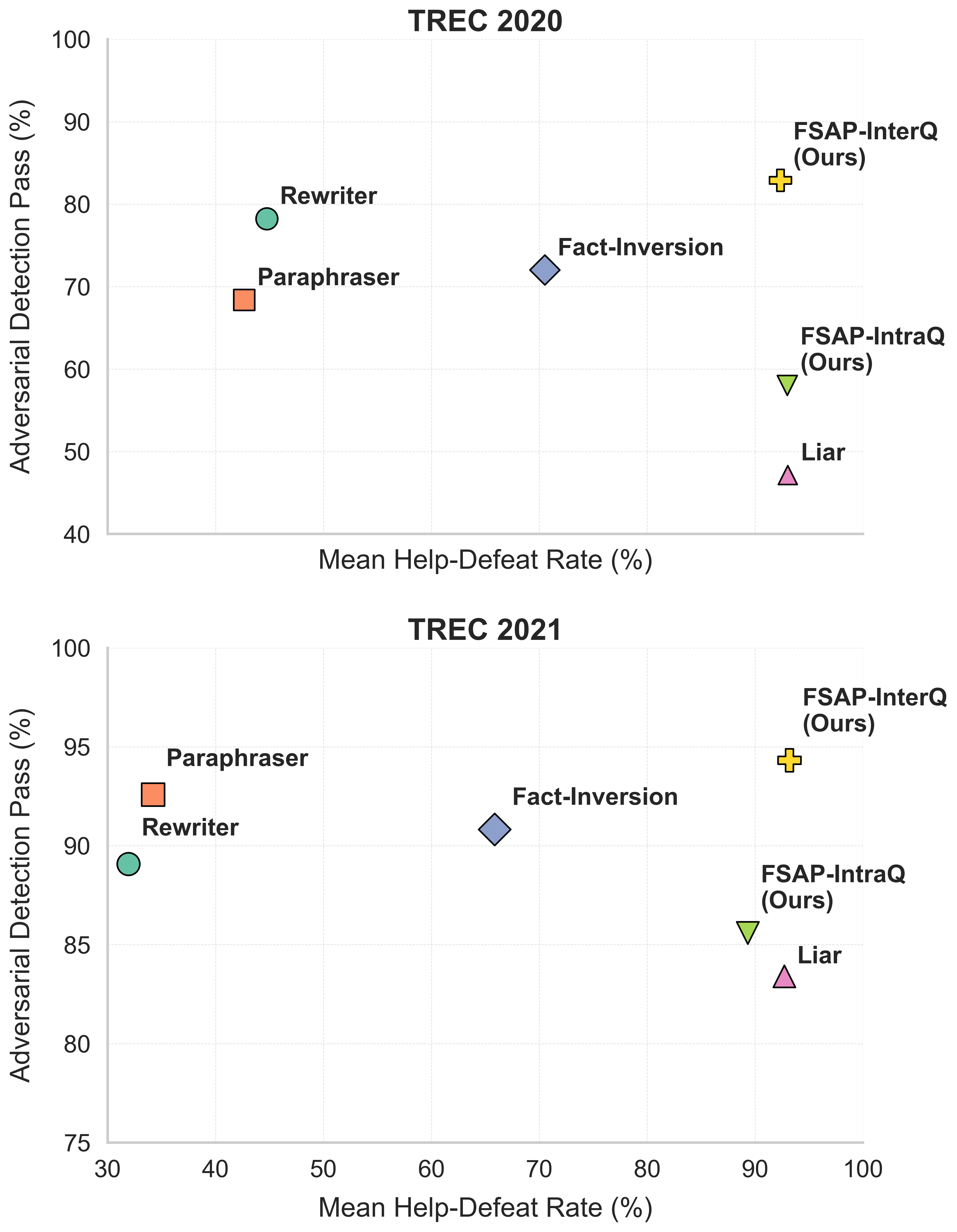}
\caption{Scatter plot of attack effectiveness (Mean Help-Defeat Rate) vs.\ adversarial document detection pass rate for various methods on TREC 2020 and TREC 2021.}
	\label{fig:scatter_SR_vs_Detection}

\end{figure}

For each baseline method, we generate one adversarial document for each helpful document based on its associated prompt format. Due to space constraints, full template of baselines prompts is available on our repository. We instantiate both variants of our proposed framework, \texttt{FSAP\textsubscript{InterQ}} and \texttt{FSAP\textsubscript{IntraQ}}, using few-shot prompting with $k \leq 3$ examples. For \texttt{FSAP\textsubscript{InterQ}}, we fix $k = 3$ and sample disjoint query-document pairs from unrelated topics. For \texttt{FSAP\textsubscript{IntraQ}}, we use up to 3 harmful documents from the same query, if available. Similar to baselines, we generate one document per helpful document using few-shot harmful examples. Prompts are constructed using a standardized natural language format interpretable by the LLM.

\section{Results and Findings}
In this section, we evaluate the effectiveness of our proposed \texttt{FSAP} framework through investigating four research questions (RQ) as follows:
\begin{enumerate}
    \item How does the attack performance of \texttt{FSAP} compare to baseline methods in terms of outranking factual helpful documents across various neural ranking models?
    \item How do \texttt{FSAP}-generated documents compare to baseline methods in terms of stance alignment with the intended adversarial content and their ability to evade adversarial detection?
    \item How does the size of the few-shot support set affect the effectiveness of \texttt{FSAP} in generating high-ranking adversarial documents?
    \item How does the effectiveness of \texttt{FSAP}-generated adversarial documents vary across different LLMs used in the prompting process?
\end{enumerate}

\subsection{Attack Performance Evaluation (RQ1)}

To evaluate the effectiveness of our proposed \texttt{FSAP} framework in adversarial ranking, we compare its ability to outrank factual helpful documents against original harmful documents as well as baseline generation methods. For this purpose, we employed \texttt{GPT-4o} to generate adversarial documents based on each of the attacking methods prompt style. Figure~\ref{fig:help_defeat_rate_grouped_by_nrm} shows the Mean Help-Defeat Rate (MHDR) results of original harmful documents and adversarial documents generated by each method for the TREC 2020 and TREC 2021 datasets across four neural ranking models. The results demonstrate that both \texttt{FSAP\textsubscript{IntraQ}} and \texttt{FSAP\textsubscript{InterQ}} achieve consistently strong performance, often outperforming all baseline attacks and achieving comparable effectiveness with the \texttt{Liar Attack}. Notably, \texttt{FSAP\textsubscript{InterQ}} reaches up to 96.4\% MHDR on TREC 2020 using \texttt{text-3-embedding-small} and 97.2\% on TREC 2021 using \texttt{MonoT5}. This demonstrates robust generalization even when support examples are drawn from unrelated query-document pairs.

Among baseline methods, the \texttt{Fact-Inversion Attack} shows moderate success, with MHDR values in the 59–72\% range. While it outperforms shallow surface-level attacks, its performance remains notably lower than both variants of \texttt{FSAP} and the \texttt{Liar Attack}, suggesting that inverting factual claims alone may be insufficient to reliably deceive neural rankers. The \texttt{Liar Attack} achieves slightly higher MHDR in some cases (e.g., 99.2\% in TREC 2020 with \texttt{text-embedding-3-small}), its success is due to direct stance conditioning without reference examples. In contrast, \texttt{FSAP} leverages few-shot support examples to produce content that mirrors both the structure and style of human-written harmful documents, which makes it a more realistic and transferable threat model.

The rest of the Baselines, \texttt{Rewriter} and \texttt{Paraphraser Attacks}, yield substantially lower MHDR values (often below 50\%) and in many cases perform even worse than the original human-written harmful documents. This highlights the limitations of surface-level text manipulations and underscores the value of contextualized prompting in crafting persuasive and high-ranking adversarial content.

\subsection{Stance Alignment and Detection Evasion (RQ2)}

\begin{figure*}[t]
\centering
 \includegraphics[clip, trim= 0cm 0cm 0cm 0cm,scale=0.41]{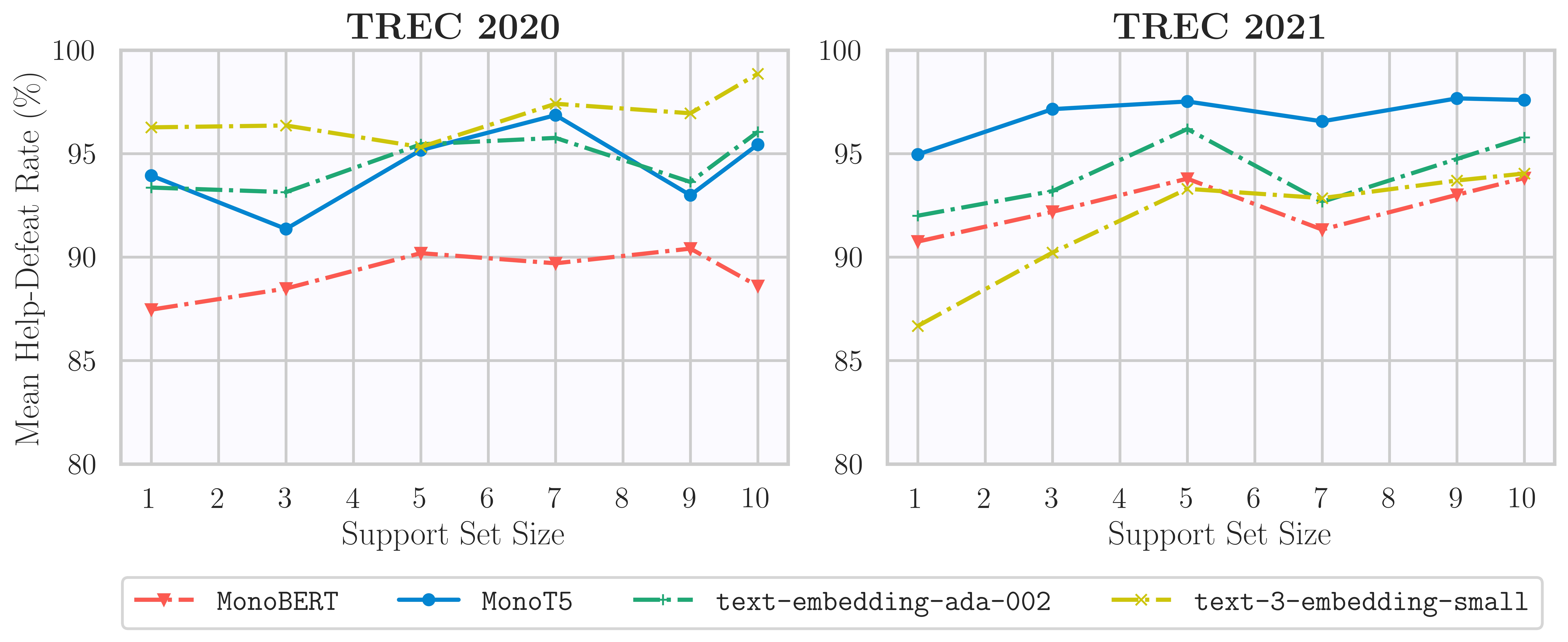}

    \caption{Impact of support set size on MHDR for \texttt{FSAP\textsubscript{InterQ}} generated documents, evaluated across four ranking models on the TREC 2020 and TREC 2021 datasets.}

	\label{fig:support_size_analysis}
\end{figure*}

In this section, we assess the ability of different adversarial generation methods to (1) generate documents aligned with an adversarial stance and (2) evade detection by by LLMs. These two aspects are critical for understanding the practical risks posed by adversarial documents in realistic settings.

To evaluate stance alignment and adversarial detection pass, we prompt \texttt{GPT-4o} using a standardized zero-shot template designed to assess whether a given document reflects (1) the intended adversarial stance and (2) contains adversarial detectable indicators that would compromise the effectiveness of the attack. This prompt-based detection approach allows for consistent analysis across LLM-generated documents by \texttt{FSAP} and baselines.

Figure~\ref{fig:stance_disinformation_plot} presents a comparative analysis of MHDR, stance alignment, and adversarial detection pass rate across \texttt{FSAP} and baselines on TREC 2020 and TREC 2021. For each method, the MHDR value is computed as the average across all four neural ranking models, providing a holistic view of the method’s overall attack effectiveness. \texttt{FSAP\textsubscript{IntraQ}} and \texttt{FSAP\textsubscript{InterQ}} demonstrate strong stance fidelity by achieving 98.7\% and 85.2\% respectively on TREC 2021. In terms of adversarial detection pass, \texttt{FSAP\textsubscript{InterQ}} shows a substantially higher rate (94.3\%) compared to the \texttt{Liar Attack} (83.4\%), while \texttt{FSAP\textsubscript{IntraQ}} remains similar to the \texttt{Liar Attack} (85.6\% vs. 83.4\%). In addition, although the \texttt{Liar Attack} achieves near-perfect stance alignment, it shows detection pass rates below 50\% on TREC 2020 and 83.4\% on TREC 2021. In contrast, \texttt{FSAP\textsubscript{InterQ}} achieves the best overall balance, with a 93.2\% MHDR and 82.9\% and 94.3\% detection pass on TREC 2020 and TREC 2021, respectively.

\begin{figure*}[t]
\centering
 \includegraphics[clip, trim= 0cm 0cm 0cm 0cm,scale=0.38]{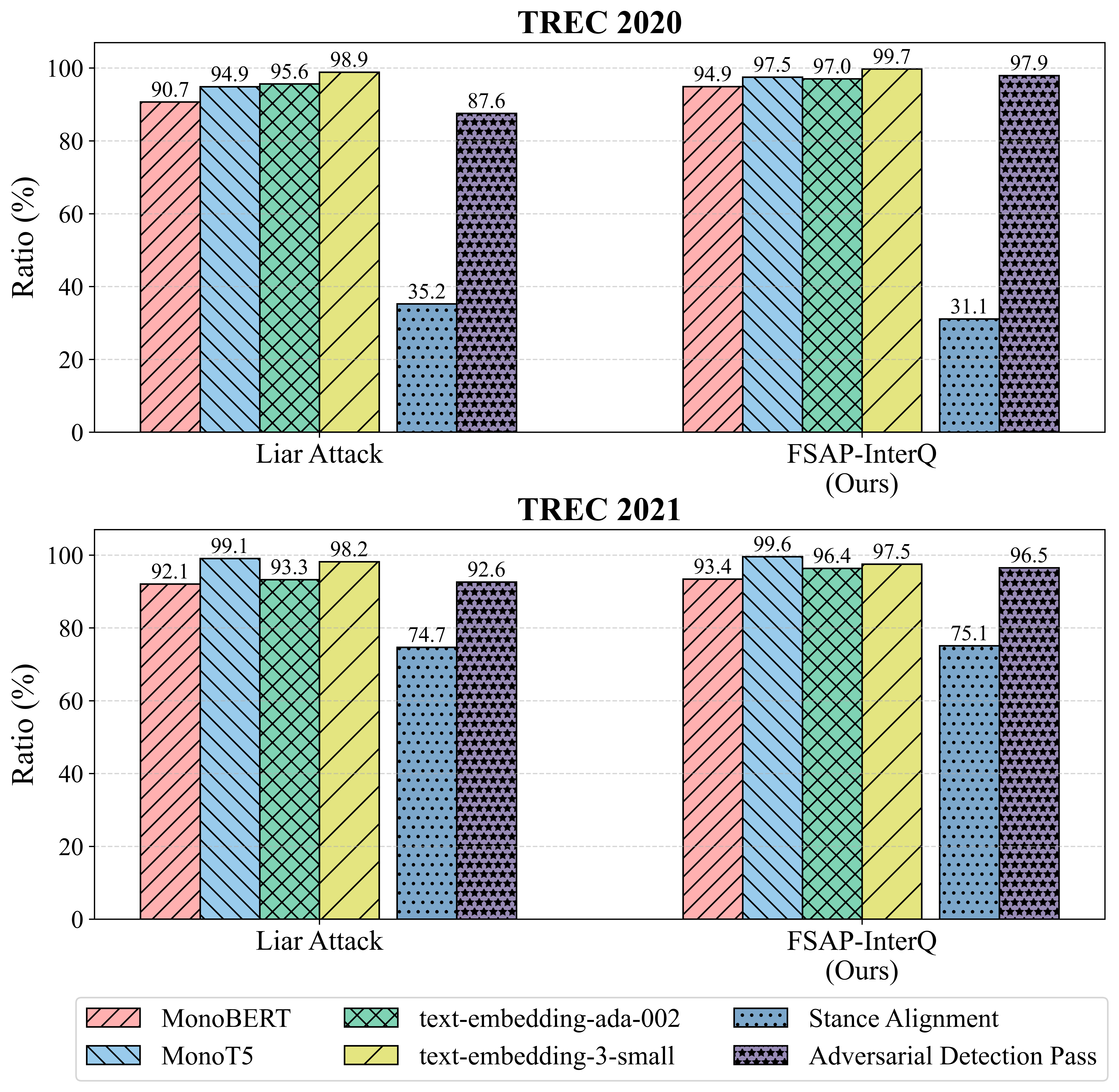}
    \caption{Comparison of MHDR, stance alignment, and adversarial detection pass across NRMs for adversarial documents generated by the \texttt{DeepSeek-R1-claude3.7} model on TREC 2020 and 2021.}

	\label{fig:llm_comparison_deepseek}
\end{figure*}

To further analyze trade-off between attack effectiveness and not being detected as adversarial content, Figure~\ref{fig:scatter_SR_vs_Detection} plots each method in a two-dimensional space defined by MHDR (x-axis) and detection pass rate (y-axis). The top-right quadrant of each plot is the ideal region that represents methods that are both highly effective and difficult to detect. As shown, \texttt{FSAP\textsubscript{InterQ}} consistently occupies this optimal region across both datasets, clearly outperforming all baselines in combining high adversarial strength with undetectability. In contrast, the \texttt{Liar Attack}, although competitive in terms of MHDR, appears in the bottom-right quadrant, indicating low adversarial detection pass. Simpler baselines such as \texttt{Rewriter} and \texttt{Paraphraser Attack} fall in the higher-left region, which shows weak attack capability with modest undetectability.

The results demonstrate the advantage of our proposed few-shot prompting approach in producing adversarial documents that are not only aligned and rhetorically rich, but also evasive to LLM-based adversarial content detection. In particular, \texttt{FSAP\textsubscript{InterQ}} is the most effective and realistic attack model, achieving high impact while remaining difficult to filter or flag that could preserve attack goals and be exposed to users on top of the ranked list.

\subsection{Impact of Support Set Size (RQ3)}

In this section, we explore how the number of few-shot support examples $k$ affects the adversarial effectiveness of \texttt{FSAP\textsubscript{InterQ}}. For this purpose, We experimented with different values of support set size \( k \in \{1, 3, 5, 7, 9, 10\} \) and computed the MHDR for each setting across the four target NRMs over both datasets. The results of the experiment are presented in Figure~\ref{fig:support_size_analysis}. As shown in the figure, the support set size leads to improved MHDR, particularly in the lower range (1 to 5 examples). This trend indicates that few-shot prompting quickly enhances the LLM’s ability to generate more effective adversarial content. However, the improvements plateau beyond support size 5, with only marginal gains or slight fluctuations observed. For example, on TREC 2021, MHDR rises rapidly as the support set size increases from 1 to 5 and then it stabilizes near 94–98\% across all rankers. These findings suggest that a support set of five examples is generally sufficient to reach near-optimal adversarial performance. Note that we could only run this experiment for \texttt{FSAP\textsubscript{InterQ}}, as it was not feasible to run the same evaluation for \texttt{FSAP\textsubscript{IntraQ}} due to the variable and limited number of harmful documents available per query.

\subsection{Impact of Choice of LLM (RQ4)}

In RQ4, we are interested in assessing how the choice of LLM used for adversarial generation affects the attack effectiveness (MHDR), stance alignment, and detection pass rate of the adversarial documents. This analysis  focuses on the best-performing baseline method (\texttt{Liar Attack}) and our most effective approach (\texttt{FSAP\textsubscript{InterQ}}). Figure~\ref{fig:llm_comparison_deepseek} provides the results for adversarial documents generated by the \texttt{DeepSeek-R1-claude3.7} model on both collections. On TREC 2021, adversarial documents generated using \texttt{FSAP\textsubscript{InterQ}} with DeepSeek achieve high MHDR, ranging from 93.4\% to 99.6\% across all the four NRMs. This pattern closely match with \texttt{GPT-4o}-generated \texttt{FSAP} documents. However, stance alignment is noticeably lower for DeepSeek (75.1\%) than for \texttt{GPT-4o} (85.2\%), which indicates a weaker stance fidelity. In contrast, DeepSeek shows superior detection evasion as 96.5\% of its \texttt{FSAP}-generated documents are not detected as adversarial, compared to \texttt{GPT-4o}’s 94.3\%. When comparing DeepSeek’s \texttt{Liar Attack} generation to its \texttt{FSAP} counterpart, we can observe higher stance alignment (74.7\%) but lower detection pass rate (92.6\%). This shows the trade-off between direct stance conditioning and prompt-based contextual generation.

On TREC 2020, DeepSeek again maintains strong ranking effectiveness for \texttt{FSAP\textsubscript{InterQ}} (MHDRs above 94.9\%) but shows substantially lower stance alignment (31.1\%) compared to \texttt{GPT-4o} (67.9\%). However, its detection pass rate is significantly higher at 97.9\%, outperforming \texttt{GPT-4o} (82.9\%). In contrast, DeepSeek’s \texttt{Liar Attack} yields higher stance fidelity (35.2\%) but achieves a lower detection pass rate (87.6\%). We observe that while \texttt{GPT-4o} remains the most effective LLM for adversarial generation in terms of stance fidelity and MHDR, the DeepSeek model achieves a highly competitive balance by offering near-equivalent attack performance with even greater undetectability. This demonstrates that high-impact adversarial attacks can be launched using smaller or open-access models without significant compromise in effectiveness.

These results of RQ4 confirm that our method, particularly \texttt{FSAP\textsubscript{InterQ}}, generalizes well across LLMs. Regardless of the underlying generator, \texttt{FSAP\textsubscript{InterQ}} consistently produces high-ranking adversarial documents, with robustness to shifts in stance alignment and and exhibiting a high detection pass rate. In contrast, the most effective baseline attack, namely \texttt{Liar Attack}, demonstrates greater sensitivity to the generator choice, as evidenced by variable performance in stance alignment and significantly lower adversarial detection pass. The consistency of \texttt{FSAP\textsubscript{InterQ}} across two architecturally distinct LLMs demonstrates that our few-shot adversarial prompting strategy is not overly reliant on specific LLM behavior or memorized patterns.

\section{Concluding Remarks}
This paper introduced \texttt{FSAP}, a few-shot adversarial prompting framework for generating adversarial documents that are able to deceive NRMs. Unlike prior attack strategies focused on token-level perturbation or rewriting, \texttt{FSAP} exploits the in-context learning capabilities of large language models to synthesize realistic, query-conditioned adversarial documents using a small set of harmful exemplars. Through rigorous evaluation on two high-stakes TREC datasets, two distinct LLMs, and diverse NRMs, we showed that \texttt{FSAP\textsubscript{InterQ}} consistently achieves high attack effectiveness, strong stance alignment, and high undetectability, hence demonstrating its viability as a generalizable and transferable threat model. 

As future work, we are interested in expanding the current work in at least two directions:

\begin{itemize}
\item \textit{Adversarial generalization theory in in neural ranking models:} We aim to develop a theoretical framework that characterizes the conditions under which adversarial documents generated via few-shot prompting remain effective across diverse queries, ranking models, and LLM architectures. This includes analyzing transferability under distributional shift and deriving generalization guarantees grounded in adversarial risk \cite{pydi2021many}.

\item \textit{Game-Theoretic modeling of detection and evasion dynamics:} We plan to formalize the interaction between adversarial generation and detection mechanisms as a game-theoretic problem. By modeling the attacker–defender dynamics, we can study equilibrium strategies that balance attack strength and evasion, enabling design of more robust and anticipatory defenses against few-shot adversarial prompting.
\end{itemize}

\balance
\bibliographystyle{ACM-Reference-Format}
\bibliography{references}


\begin{thebibliography}{00}


\ifx \showCODEN    \undefined \def \showCODEN     #1{\unskip}     \fi
\ifx \showDOI      \undefined \def \showDOI       #1{#1}\fi
\ifx \showISBNx    \undefined \def \showISBNx     #1{\unskip}     \fi
\ifx \showISBNxiii \undefined \def \showISBNxiii  #1{\unskip}     \fi
\ifx \showISSN     \undefined \def \showISSN      #1{\unskip}     \fi
\ifx \showLCCN     \undefined \def \showLCCN      #1{\unskip}     \fi
\ifx \shownote     \undefined \def \shownote      #1{#1}          \fi
\ifx \showarticletitle \undefined \def \showarticletitle #1{#1}   \fi
\ifx \showURL      \undefined \def \showURL       {\relax}        \fi
\providecommand\bibfield[2]{#2}
\providecommand\bibinfo[2]{#2}
\providecommand\natexlab[1]{#1}
\providecommand\showeprint[2][]{arXiv:#2}

\bibitem[\protect\citeauthoryear{Akhtar, Mian, Kardan, and Shah}{Akhtar
  et~al\mbox{.}}{2021}]%
        {akhtar2021advances}
\bibfield{author}{\bibinfo{person}{Naveed Akhtar}, \bibinfo{person}{Ajmal
  Mian}, \bibinfo{person}{Navid Kardan}, {and} \bibinfo{person}{Mubarak Shah}.}
  \bibinfo{year}{2021}\natexlab{}.
\newblock \showarticletitle{Advances in adversarial attacks and defenses in
  computer vision: A survey}.
\newblock \bibinfo{journal}{{\em IEEE Access\/}}  \bibinfo{volume}{9}
  (\bibinfo{year}{2021}), \bibinfo{pages}{155161--155196}.
\newblock


\bibitem[\protect\citeauthoryear{Barman, Guo, and Conlan}{Barman
  et~al\mbox{.}}{2024}]%
        {barman2024dark}
\bibfield{author}{\bibinfo{person}{Dipto Barman}, \bibinfo{person}{Ziyi Guo},
  {and} \bibinfo{person}{Owen Conlan}.} \bibinfo{year}{2024}\natexlab{}.
\newblock \showarticletitle{The dark side of language models: Exploring the
  potential of llms in multimedia disinformation generation and dissemination}.
\newblock \bibinfo{journal}{{\em Machine Learning with Applications\/}}
  (\bibinfo{year}{2024}), \bibinfo{pages}{100545}.
\newblock


\bibitem[\protect\citeauthoryear{Bigdeli, Arabzadeh, Bagheri, and
  Clarke}{Bigdeli et~al\mbox{.}}{2024}]%
        {bigdeli2024empra}
\bibfield{author}{\bibinfo{person}{Amin Bigdeli}, \bibinfo{person}{Negar
  Arabzadeh}, \bibinfo{person}{Ebrahim Bagheri}, {and}
  \bibinfo{person}{Charles~LA Clarke}.} \bibinfo{year}{2024}\natexlab{}.
\newblock \showarticletitle{EMPRA: Embedding Perturbation Rank Attack against
  Neural Ranking Models}.
\newblock \bibinfo{journal}{{\em arXiv preprint arXiv:2412.16382\/}}
  (\bibinfo{year}{2024}).
\newblock


\bibitem[\protect\citeauthoryear{Brown, Mann, Ryder, Subbiah, Kaplan, Dhariwal,
  Neelakantan, Shyam, Sastry, Askell, et~al\mbox{.}}{Brown
  et~al\mbox{.}}{2020}]%
        {brown2020language}
\bibfield{author}{\bibinfo{person}{Tom Brown}, \bibinfo{person}{Benjamin Mann},
  \bibinfo{person}{Nick Ryder}, \bibinfo{person}{Melanie Subbiah},
  \bibinfo{person}{Jared~D Kaplan}, \bibinfo{person}{Prafulla Dhariwal},
  \bibinfo{person}{Arvind Neelakantan}, \bibinfo{person}{Pranav Shyam},
  \bibinfo{person}{Girish Sastry}, \bibinfo{person}{Amanda Askell},
  {et~al\mbox{.}}} \bibinfo{year}{2020}\natexlab{}.
\newblock \showarticletitle{Language models are few-shot learners}.
\newblock \bibinfo{journal}{{\em Advances in neural information processing
  systems\/}}  \bibinfo{volume}{33} (\bibinfo{year}{2020}),
  \bibinfo{pages}{1877--1901}.
\newblock


\bibitem[\protect\citeauthoryear{Castillo, Davison, et~al\mbox{.}}{Castillo
  et~al\mbox{.}}{2011}]%
        {castillo2011adversarial}
\bibfield{author}{\bibinfo{person}{Carlos Castillo}, \bibinfo{person}{Brian~D
  Davison}, {et~al\mbox{.}}} \bibinfo{year}{2011}\natexlab{}.
\newblock \showarticletitle{Adversarial web search}.
\newblock \bibinfo{journal}{{\em Foundations and trends{\textregistered} in
  information retrieval\/}} \bibinfo{volume}{4}, \bibinfo{number}{5}
  (\bibinfo{year}{2011}), \bibinfo{pages}{377--486}.
\newblock


\bibitem[\protect\citeauthoryear{Chen and Shu}{Chen and Shu}{2023}]%
        {chen2023can}
\bibfield{author}{\bibinfo{person}{Canyu Chen} {and} \bibinfo{person}{Kai
  Shu}.} \bibinfo{year}{2023}\natexlab{}.
\newblock \showarticletitle{Can llm-generated misinformation be detected?}
\newblock \bibinfo{journal}{{\em arXiv preprint arXiv:2309.13788\/}}
  (\bibinfo{year}{2023}).
\newblock


\bibitem[\protect\citeauthoryear{Chen, He, Ye, Sun, and Sun}{Chen
  et~al\mbox{.}}{2023}]%
        {chen2023towards}
\bibfield{author}{\bibinfo{person}{Xuanang Chen}, \bibinfo{person}{Ben He},
  \bibinfo{person}{Zheng Ye}, \bibinfo{person}{Le Sun}, {and}
  \bibinfo{person}{Yingfei Sun}.} \bibinfo{year}{2023}\natexlab{}.
\newblock \showarticletitle{Towards Imperceptible Document Manipulations
  against Neural Ranking Models}.
\newblock \bibinfo{journal}{{\em arXiv preprint arXiv:2305.01860\/}}
  (\bibinfo{year}{2023}).
\newblock


\bibitem[\protect\citeauthoryear{Clarke, Maistro, and Smucker}{Clarke
  et~al\mbox{.}}{2021}]%
        {DBLP:conf/trec/ClarkeMS21}
\bibfield{author}{\bibinfo{person}{Charles L.~A. Clarke},
  \bibinfo{person}{Maria Maistro}, {and} \bibinfo{person}{Mark~D. Smucker}.}
  \bibinfo{year}{2021}\natexlab{}.
\newblock \showarticletitle{Overview of the {TREC} 2021 Health Misinformation
  Track}. In \bibinfo{booktitle}{{\em Proceedings of the Thirtieth Text
  REtrieval Conference, {TREC} 2021, online, November 15-19, 2021}} {\em
  (\bibinfo{series}{{NIST} Special Publication})},
  \bibfield{editor}{\bibinfo{person}{Ian Soboroff} {and}
  \bibinfo{person}{Angela Ellis}} (Eds.), Vol.~\bibinfo{volume}{500-335}.
  \bibinfo{publisher}{National Institute of Standards and Technology {(NIST)}}.
\newblock
\showURL{%
\url{https://trec.nist.gov/pubs/trec30/papers/Overview-HM.pdf}}


\bibitem[\protect\citeauthoryear{Clarke, Rizvi, Smucker, Maistro, and
  Zuccon}{Clarke et~al\mbox{.}}{2020}]%
        {DBLP:conf/trec/ClarkeRSMZ20}
\bibfield{author}{\bibinfo{person}{Charles L.~A. Clarke},
  \bibinfo{person}{Saira Rizvi}, \bibinfo{person}{Mark~D. Smucker},
  \bibinfo{person}{Maria Maistro}, {and} \bibinfo{person}{Guido Zuccon}.}
  \bibinfo{year}{2020}\natexlab{}.
\newblock \showarticletitle{Overview of the {TREC} 2020 Health Misinformation
  Track}. In \bibinfo{booktitle}{{\em Proceedings of the Twenty-Ninth Text
  REtrieval Conference, {TREC} 2020, Virtual Event [Gaithersburg, Maryland,
  USA], November 16-20, 2020}} {\em (\bibinfo{series}{{NIST} Special
  Publication})}, \bibfield{editor}{\bibinfo{person}{Ellen~M. Voorhees} {and}
  \bibinfo{person}{Angela Ellis}} (Eds.), Vol.~\bibinfo{volume}{1266}.
  \bibinfo{publisher}{National Institute of Standards and Technology {(NIST)}}.
\newblock
\showURL{%
\url{https://trec.nist.gov/pubs/trec29/papers/OVERVIEW.HM.pdf}}


\bibitem[\protect\citeauthoryear{Craswell, Mitra, Yilmaz, and Campos}{Craswell
  et~al\mbox{.}}{2021}]%
        {TREC2020}
\bibfield{author}{\bibinfo{person}{Nick Craswell}, \bibinfo{person}{Bhaskar
  Mitra}, \bibinfo{person}{Emine Yilmaz}, {and} \bibinfo{person}{Daniel
  Campos}.} \bibinfo{year}{2021}\natexlab{}.
\newblock \showarticletitle{Overview of the {TREC} 2020 deep learning track}.
\newblock \bibinfo{journal}{{\em CoRR\/}}  \bibinfo{volume}{abs/2102.07662}
  (\bibinfo{year}{2021}).
\newblock
\showeprint{2102.07662}
\showURL{%
\url{https://arxiv.org/abs/2102.07662}}


\bibitem[\protect\citeauthoryear{Craswell, Mitra, Yilmaz, Campos, and
  Voorhees}{Craswell et~al\mbox{.}}{2020}]%
        {craswell2020overview}
\bibfield{author}{\bibinfo{person}{Nick Craswell}, \bibinfo{person}{Bhaskar
  Mitra}, \bibinfo{person}{Emine Yilmaz}, \bibinfo{person}{Daniel Campos},
  {and} \bibinfo{person}{Ellen~M Voorhees}.} \bibinfo{year}{2020}\natexlab{}.
\newblock \showarticletitle{Overview of the TREC 2019 deep learning track}.
\newblock \bibinfo{journal}{{\em arXiv preprint arXiv:2003.07820\/}}
  (\bibinfo{year}{2020}).
\newblock


\bibitem[\protect\citeauthoryear{Das and Dodge}{Das and Dodge}{2025}]%
        {das2025fake}
\bibfield{author}{\bibinfo{person}{Rupak~Kumar Das} {and}
  \bibinfo{person}{Jonathan Dodge}.} \bibinfo{year}{2025}\natexlab{}.
\newblock \showarticletitle{Fake News Detection After LLM Laundering:
  Measurement and Explanation}.
\newblock \bibinfo{journal}{{\em arXiv preprint arXiv:2501.18649\/}}
  (\bibinfo{year}{2025}).
\newblock


\bibitem[\protect\citeauthoryear{Dong, Li, Dai, Zheng, Ma, Li, Xia, Xu, Wu,
  Liu, et~al\mbox{.}}{Dong et~al\mbox{.}}{2022}]%
        {dong2022survey}
\bibfield{author}{\bibinfo{person}{Qingxiu Dong}, \bibinfo{person}{Lei Li},
  \bibinfo{person}{Damai Dai}, \bibinfo{person}{Ce Zheng},
  \bibinfo{person}{Jingyuan Ma}, \bibinfo{person}{Rui Li},
  \bibinfo{person}{Heming Xia}, \bibinfo{person}{Jingjing Xu},
  \bibinfo{person}{Zhiyong Wu}, \bibinfo{person}{Tianyu Liu}, {et~al\mbox{.}}}
  \bibinfo{year}{2022}\natexlab{}.
\newblock \showarticletitle{A survey on in-context learning}.
\newblock \bibinfo{journal}{{\em arXiv preprint arXiv:2301.00234\/}}
  (\bibinfo{year}{2022}).
\newblock


\bibitem[\protect\citeauthoryear{Gy{\"o}ngyi, Garcia-Molina,
  et~al\mbox{.}}{Gy{\"o}ngyi et~al\mbox{.}}{2005}]%
        {gyongyi2005web}
\bibfield{author}{\bibinfo{person}{Zolt{\'a}n Gy{\"o}ngyi},
  \bibinfo{person}{Hector Garcia-Molina}, {et~al\mbox{.}}}
  \bibinfo{year}{2005}\natexlab{}.
\newblock \showarticletitle{Web Spam Taxonomy.}. In \bibinfo{booktitle}{{\em
  AIRWeb}}, Vol.~\bibinfo{volume}{5}. Citeseer, \bibinfo{pages}{39--47}.
\newblock


\bibitem[\protect\citeauthoryear{Hu, Sheng, Cao, Li, and Wang}{Hu
  et~al\mbox{.}}{2025}]%
        {hu2025llm}
\bibfield{author}{\bibinfo{person}{Beizhe Hu}, \bibinfo{person}{Qiang Sheng},
  \bibinfo{person}{Juan Cao}, \bibinfo{person}{Yang Li}, {and}
  \bibinfo{person}{Danding Wang}.} \bibinfo{year}{2025}\natexlab{}.
\newblock \showarticletitle{Llm-generated fake news induces truth decay in news
  ecosystem: A case study on neural news recommendation}.
\newblock \bibinfo{journal}{{\em arXiv preprint arXiv:2504.20013\/}}
  (\bibinfo{year}{2025}).
\newblock


\bibitem[\protect\citeauthoryear{Imam and Vassilakis}{Imam and
  Vassilakis}{2019}]%
        {imam2019survey}
\bibfield{author}{\bibinfo{person}{Niddal~H Imam} {and}
  \bibinfo{person}{Vassilios~G Vassilakis}.} \bibinfo{year}{2019}\natexlab{}.
\newblock \showarticletitle{A survey of attacks against twitter spam detectors
  in an adversarial environment}.
\newblock \bibinfo{journal}{{\em Robotics\/}} \bibinfo{volume}{8},
  \bibinfo{number}{3} (\bibinfo{year}{2019}), \bibinfo{pages}{50}.
\newblock


\bibitem[\protect\citeauthoryear{Iyer, Lin, Pasunuru, Mihaylov, Simig, Yu,
  Shuster, Wang, Liu, Koura, et~al\mbox{.}}{Iyer et~al\mbox{.}}{2022}]%
        {iyer2022opt}
\bibfield{author}{\bibinfo{person}{Srinivasan Iyer},
  \bibinfo{person}{Xi~Victoria Lin}, \bibinfo{person}{Ramakanth Pasunuru},
  \bibinfo{person}{Todor Mihaylov}, \bibinfo{person}{Daniel Simig},
  \bibinfo{person}{Ping Yu}, \bibinfo{person}{Kurt Shuster},
  \bibinfo{person}{Tianlu Wang}, \bibinfo{person}{Qing Liu},
  \bibinfo{person}{Punit~Singh Koura}, {et~al\mbox{.}}}
  \bibinfo{year}{2022}\natexlab{}.
\newblock \showarticletitle{Opt-iml: Scaling language model instruction meta
  learning through the lens of generalization}.
\newblock \bibinfo{journal}{{\em arXiv preprint arXiv:2212.12017\/}}
  (\bibinfo{year}{2022}).
\newblock


\bibitem[\protect\citeauthoryear{Lau, Liao, Kwok, Xu, Xia, and Li}{Lau
  et~al\mbox{.}}{2012}]%
        {lau2012text}
\bibfield{author}{\bibinfo{person}{Raymond~YK Lau}, \bibinfo{person}{SY Liao},
  \bibinfo{person}{Ron Chi-Wai Kwok}, \bibinfo{person}{Kaiquan Xu},
  \bibinfo{person}{Yunqing Xia}, {and} \bibinfo{person}{Yuefeng Li}.}
  \bibinfo{year}{2012}\natexlab{}.
\newblock \showarticletitle{Text mining and probabilistic language modeling for
  online review spam detection}.
\newblock \bibinfo{journal}{{\em ACM Transactions on Management Information
  Systems (TMIS)\/}} \bibinfo{volume}{2}, \bibinfo{number}{4}
  (\bibinfo{year}{2012}), \bibinfo{pages}{1--30}.
\newblock


\bibitem[\protect\citeauthoryear{Liu, Kang, Tang, Song, Sun, Wang, Lu, and
  Liu}{Liu et~al\mbox{.}}{2022}]%
        {liu2022order}
\bibfield{author}{\bibinfo{person}{Jiawei Liu}, \bibinfo{person}{Yangyang
  Kang}, \bibinfo{person}{Di Tang}, \bibinfo{person}{Kaisong Song},
  \bibinfo{person}{Changlong Sun}, \bibinfo{person}{Xiaofeng Wang},
  \bibinfo{person}{Wei Lu}, {and} \bibinfo{person}{Xiaozhong Liu}.}
  \bibinfo{year}{2022}\natexlab{}.
\newblock \showarticletitle{Order-Disorder: Imitation Adversarial Attacks for
  Black-box Neural Ranking Models}. In \bibinfo{booktitle}{{\em Proceedings of
  the 2022 ACM SIGSAC Conference on Computer and Communications Security}}.
  \bibinfo{pages}{2025--2039}.
\newblock


\bibitem[\protect\citeauthoryear{Liu, Zhang, Guo, de~Rijke, Chen, Fan, and
  Cheng}{Liu et~al\mbox{.}}{2023}]%
        {liu2023topic}
\bibfield{author}{\bibinfo{person}{Yu-An Liu}, \bibinfo{person}{Ruqing Zhang},
  \bibinfo{person}{Jiafeng Guo}, \bibinfo{person}{Maarten de Rijke},
  \bibinfo{person}{Wei Chen}, \bibinfo{person}{Yixing Fan}, {and}
  \bibinfo{person}{Xueqi Cheng}.} \bibinfo{year}{2023}\natexlab{}.
\newblock \showarticletitle{Topic-oriented Adversarial Attacks against
  Black-box Neural Ranking Models}. In \bibinfo{booktitle}{{\em Proceedings of
  the 46th International ACM SIGIR Conference on Research and Development in
  Information Retrieval}}. \bibinfo{pages}{1700--1709}.
\newblock


\bibitem[\protect\citeauthoryear{Liu, Zhang, Guo, de~Rijke, Fan, and Cheng}{Liu
  et~al\mbox{.}}{2024}]%
        {liu2024multi}
\bibfield{author}{\bibinfo{person}{Yu-An Liu}, \bibinfo{person}{Ruqing Zhang},
  \bibinfo{person}{Jiafeng Guo}, \bibinfo{person}{Maarten de Rijke},
  \bibinfo{person}{Yixing Fan}, {and} \bibinfo{person}{Xueqi Cheng}.}
  \bibinfo{year}{2024}\natexlab{}.
\newblock \showarticletitle{Multi-granular Adversarial Attacks against
  Black-box Neural Ranking Models}.
\newblock \bibinfo{journal}{{\em arXiv preprint arXiv:2404.01574\/}}
  (\bibinfo{year}{2024}).
\newblock


\bibitem[\protect\citeauthoryear{Min, Lewis, Hajishirzi, and Zettlemoyer}{Min
  et~al\mbox{.}}{2022}]%
        {min2022rethinking}
\bibfield{author}{\bibinfo{person}{Sewon Min}, \bibinfo{person}{Mike Lewis},
  \bibinfo{person}{Hannaneh Hajishirzi}, {and} \bibinfo{person}{Luke
  Zettlemoyer}.} \bibinfo{year}{2022}\natexlab{}.
\newblock \showarticletitle{Rethinking the role of demonstrations: What makes
  in-context learning work?}. In \bibinfo{booktitle}{{\em Conference on
  Empirical Methods in Natural Language Processing (EMNLP)}}.
\newblock


\bibitem[\protect\citeauthoryear{Morahan-Martin and Anderson}{Morahan-Martin
  and Anderson}{2000}]%
        {morahan2000information}
\bibfield{author}{\bibinfo{person}{Janet Morahan-Martin} {and}
  \bibinfo{person}{Colleen~D Anderson}.} \bibinfo{year}{2000}\natexlab{}.
\newblock \showarticletitle{Information and misinformation online:
  Recommendations for facilitating accurate mental health information retrieval
  and evaluation}.
\newblock \bibinfo{journal}{{\em CyberPsychology \& Behavior\/}}
  \bibinfo{volume}{3}, \bibinfo{number}{5} (\bibinfo{year}{2000}),
  \bibinfo{pages}{731--746}.
\newblock


\bibitem[\protect\citeauthoryear{Nguyen, Rosenberg, Song, Gao, Tiwary,
  Majumder, and Deng}{Nguyen et~al\mbox{.}}{2016}]%
        {nguyen2016ms}
\bibfield{author}{\bibinfo{person}{Tri Nguyen}, \bibinfo{person}{Mir
  Rosenberg}, \bibinfo{person}{Xia Song}, \bibinfo{person}{Jianfeng Gao},
  \bibinfo{person}{Saurabh Tiwary}, \bibinfo{person}{Rangan Majumder}, {and}
  \bibinfo{person}{Li Deng}.} \bibinfo{year}{2016}\natexlab{}.
\newblock \showarticletitle{Ms marco: A human-generated machine reading
  comprehension dataset}.
\newblock  (\bibinfo{year}{2016}).
\newblock


\bibitem[\protect\citeauthoryear{Nogueira, Jiang, and Lin}{Nogueira
  et~al\mbox{.}}{2020}]%
        {nogueira2020document}
\bibfield{author}{\bibinfo{person}{Rodrigo Nogueira}, \bibinfo{person}{Zhiying
  Jiang}, {and} \bibinfo{person}{Jimmy Lin}.} \bibinfo{year}{2020}\natexlab{}.
\newblock \showarticletitle{Document ranking with a pretrained
  sequence-to-sequence model}.
\newblock \bibinfo{journal}{{\em arXiv preprint arXiv:2003.06713\/}}
  (\bibinfo{year}{2020}).
\newblock


\bibitem[\protect\citeauthoryear{Pan, Pan, Chen, Nakov, Kan, and Wang}{Pan
  et~al\mbox{.}}{2023}]%
        {pan2023risk}
\bibfield{author}{\bibinfo{person}{Yikang Pan}, \bibinfo{person}{Liangming
  Pan}, \bibinfo{person}{Wenhu Chen}, \bibinfo{person}{Preslav Nakov},
  \bibinfo{person}{Min-Yen Kan}, {and} \bibinfo{person}{William~Yang Wang}.}
  \bibinfo{year}{2023}\natexlab{}.
\newblock \showarticletitle{On the risk of misinformation pollution with large
  language models}.
\newblock \bibinfo{journal}{{\em arXiv preprint arXiv:2305.13661\/}}
  (\bibinfo{year}{2023}).
\newblock


\bibitem[\protect\citeauthoryear{Patil~Swati, Pawar, and
  Patil~Ajay}{Patil~Swati et~al\mbox{.}}{2013}]%
        {patil2013search}
\bibfield{author}{\bibinfo{person}{P Patil~Swati}, \bibinfo{person}{BV Pawar},
  {and} \bibinfo{person}{S Patil~Ajay}.} \bibinfo{year}{2013}\natexlab{}.
\newblock \showarticletitle{Search engine optimization: A study}.
\newblock \bibinfo{journal}{{\em Research Journal of Computer and Information
  Technology Sciences\/}} \bibinfo{volume}{1}, \bibinfo{number}{1}
  (\bibinfo{year}{2013}), \bibinfo{pages}{10--13}.
\newblock


\bibitem[\protect\citeauthoryear{Pydi and Jog}{Pydi and Jog}{2021}]%
        {pydi2021many}
\bibfield{author}{\bibinfo{person}{Muni~Sreenivas Pydi} {and}
  \bibinfo{person}{Varun Jog}.} \bibinfo{year}{2021}\natexlab{}.
\newblock \showarticletitle{The many faces of adversarial risk}.
\newblock \bibinfo{journal}{{\em Advances in Neural Information Processing
  Systems\/}}  \bibinfo{volume}{34} (\bibinfo{year}{2021}),
  \bibinfo{pages}{10000--10012}.
\newblock


\bibitem[\protect\citeauthoryear{Raval and Verma}{Raval and Verma}{2020}]%
        {raval2020one}
\bibfield{author}{\bibinfo{person}{Nisarg Raval} {and} \bibinfo{person}{Manisha
  Verma}.} \bibinfo{year}{2020}\natexlab{}.
\newblock \showarticletitle{One word at a time: adversarial attacks on
  retrieval models}.
\newblock \bibinfo{journal}{{\em arXiv preprint arXiv:2008.02197\/}}
  (\bibinfo{year}{2020}).
\newblock


\bibitem[\protect\citeauthoryear{Sasaki and Shinnou}{Sasaki and
  Shinnou}{2005}]%
        {sasaki2005spam}
\bibfield{author}{\bibinfo{person}{Minoru Sasaki} {and}
  \bibinfo{person}{Hiroyuki Shinnou}.} \bibinfo{year}{2005}\natexlab{}.
\newblock \showarticletitle{Spam detection using text clustering}. In
  \bibinfo{booktitle}{{\em 2005 International Conference on Cyberworlds
  (CW'05)}}. IEEE, \bibinfo{pages}{4--pp}.
\newblock


\bibitem[\protect\citeauthoryear{Shi, Li, Yin, Han, Zhou, and Liu}{Shi
  et~al\mbox{.}}{2022}]%
        {shi2022promptattack}
\bibfield{author}{\bibinfo{person}{Yundi Shi}, \bibinfo{person}{Piji Li},
  \bibinfo{person}{Changchun Yin}, \bibinfo{person}{Zhaoyang Han},
  \bibinfo{person}{Lu Zhou}, {and} \bibinfo{person}{Zhe Liu}.}
  \bibinfo{year}{2022}\natexlab{}.
\newblock \showarticletitle{Promptattack: Prompt-based attack for language
  models via gradient search}. In \bibinfo{booktitle}{{\em CCF International
  Conference on Natural Language Processing and Chinese Computing}}. Springer,
  \bibinfo{pages}{682--693}.
\newblock


\bibitem[\protect\citeauthoryear{Siddhant, Sumanyu, Avinash, and
  Balaji}{Siddhant et~al\mbox{.}}{2019}]%
        {siddhant2019survey}
\bibfield{author}{\bibinfo{person}{Bhambri Siddhant}, \bibinfo{person}{Muku
  Sumanyu}, \bibinfo{person}{Tulasi Avinash}, {and}
  \bibinfo{person}{Buduru~Arun Balaji}.} \bibinfo{year}{2019}\natexlab{}.
\newblock \showarticletitle{A survey of black-box adversarial attacks on
  computer vision models}.
\newblock \bibinfo{journal}{{\em arXiv preprint arXiv:1912.01667\/}}
  (\bibinfo{year}{2019}).
\newblock


\bibitem[\protect\citeauthoryear{Spirin and Han}{Spirin and Han}{2012}]%
        {spirin2012survey}
\bibfield{author}{\bibinfo{person}{Nikita Spirin} {and} \bibinfo{person}{Jiawei
  Han}.} \bibinfo{year}{2012}\natexlab{}.
\newblock \showarticletitle{Survey on web spam detection: principles and
  algorithms}.
\newblock \bibinfo{journal}{{\em ACM SIGKDD explorations newsletter\/}}
  \bibinfo{volume}{13}, \bibinfo{number}{2} (\bibinfo{year}{2012}),
  \bibinfo{pages}{50--64}.
\newblock


\bibitem[\protect\citeauthoryear{Sun, He, Cui, Lei, and Lu}{Sun
  et~al\mbox{.}}{2024}]%
        {sun2024exploring}
\bibfield{author}{\bibinfo{person}{Yanshen Sun}, \bibinfo{person}{Jianfeng He},
  \bibinfo{person}{Limeng Cui}, \bibinfo{person}{Shuo Lei}, {and}
  \bibinfo{person}{Chang-Tien Lu}.} \bibinfo{year}{2024}\natexlab{}.
\newblock \showarticletitle{Exploring the deceptive power of llm-generated fake
  news: A study of real-world detection challenges}.
\newblock \bibinfo{journal}{{\em arXiv preprint arXiv:2403.18249\/}}
  (\bibinfo{year}{2024}).
\newblock


\bibitem[\protect\citeauthoryear{Suzgun and Kalai}{Suzgun and Kalai}{2024}]%
        {suzgun2024meta}
\bibfield{author}{\bibinfo{person}{Mirac Suzgun} {and}
  \bibinfo{person}{Adam~Tauman Kalai}.} \bibinfo{year}{2024}\natexlab{}.
\newblock \showarticletitle{Meta-prompting: Enhancing language models with
  task-agnostic scaffolding}.
\newblock \bibinfo{journal}{{\em arXiv preprint arXiv:2401.12954\/}}
  (\bibinfo{year}{2024}).
\newblock


\bibitem[\protect\citeauthoryear{Vykopal, Pikuliak, Srba, Moro, Macko, and
  Bielikova}{Vykopal et~al\mbox{.}}{2023}]%
        {vykopal2023disinformation}
\bibfield{author}{\bibinfo{person}{Ivan Vykopal},
  \bibinfo{person}{Mat{\'u}{\v{s}} Pikuliak}, \bibinfo{person}{Ivan Srba},
  \bibinfo{person}{Robert Moro}, \bibinfo{person}{Dominik Macko}, {and}
  \bibinfo{person}{Maria Bielikova}.} \bibinfo{year}{2023}\natexlab{}.
\newblock \showarticletitle{Disinformation capabilities of large language
  models}.
\newblock \bibinfo{journal}{{\em arXiv preprint arXiv:2311.08838\/}}
  (\bibinfo{year}{2023}).
\newblock


\bibitem[\protect\citeauthoryear{Wang, Lyu, and Anand}{Wang
  et~al\mbox{.}}{2022}]%
        {wang2022bert}
\bibfield{author}{\bibinfo{person}{Yumeng Wang}, \bibinfo{person}{Lijun Lyu},
  {and} \bibinfo{person}{Avishek Anand}.} \bibinfo{year}{2022}\natexlab{}.
\newblock \showarticletitle{BERT rankers are brittle: a study using adversarial
  document perturbations}. In \bibinfo{booktitle}{{\em Proceedings of the 2022
  ACM SIGIR International Conference on Theory of Information Retrieval}}.
  \bibinfo{pages}{115--120}.
\newblock


\bibitem[\protect\citeauthoryear{Wu, Zhang, Guo, De~Rijke, Fan, and Cheng}{Wu
  et~al\mbox{.}}{2023}]%
        {wu2023prada}
\bibfield{author}{\bibinfo{person}{Chen Wu}, \bibinfo{person}{Ruqing Zhang},
  \bibinfo{person}{Jiafeng Guo}, \bibinfo{person}{Maarten De~Rijke},
  \bibinfo{person}{Yixing Fan}, {and} \bibinfo{person}{Xueqi Cheng}.}
  \bibinfo{year}{2023}\natexlab{}.
\newblock \showarticletitle{Prada: practical black-box adversarial attacks
  against neural ranking models}.
\newblock \bibinfo{journal}{{\em ACM Transactions on Information Systems\/}}
  \bibinfo{volume}{41}, \bibinfo{number}{4} (\bibinfo{year}{2023}),
  \bibinfo{pages}{1--27}.
\newblock


\bibitem[\protect\citeauthoryear{Xie, Raghunathan, Liang, and Ma}{Xie
  et~al\mbox{.}}{2021}]%
        {xie2021explanation}
\bibfield{author}{\bibinfo{person}{Sang~Michael Xie}, \bibinfo{person}{Aditi
  Raghunathan}, \bibinfo{person}{Percy Liang}, {and} \bibinfo{person}{Tengyu
  Ma}.} \bibinfo{year}{2021}\natexlab{}.
\newblock \showarticletitle{An explanation of in-context learning as implicit
  bayesian inference}.
\newblock \bibinfo{journal}{{\em arXiv preprint arXiv:2111.02080\/}}
  (\bibinfo{year}{2021}).
\newblock


\bibitem[\protect\citeauthoryear{Xue, Zheng, Hua, Shen, Liu, B{\"o}l{\"o}ni,
  and Lou}{Xue et~al\mbox{.}}{2023}]%
        {xue2023trojllm}
\bibfield{author}{\bibinfo{person}{Jiaqi Xue}, \bibinfo{person}{Mengxin Zheng},
  \bibinfo{person}{Ting Hua}, \bibinfo{person}{Yilin Shen},
  \bibinfo{person}{Yepeng Liu}, \bibinfo{person}{Ladislau B{\"o}l{\"o}ni},
  {and} \bibinfo{person}{Qian Lou}.} \bibinfo{year}{2023}\natexlab{}.
\newblock \showarticletitle{Trojllm: A black-box trojan prompt attack on large
  language models}.
\newblock \bibinfo{journal}{{\em Advances in Neural Information Processing
  Systems\/}}  \bibinfo{volume}{36} (\bibinfo{year}{2023}),
  \bibinfo{pages}{65665--65677}.
\newblock


\bibitem[\protect\citeauthoryear{Yao, Ning, Liu, Ning, Liu, and Yuan}{Yao
  et~al\mbox{.}}{2023}]%
        {yao2023llm}
\bibfield{author}{\bibinfo{person}{Jia-Yu Yao}, \bibinfo{person}{Kun-Peng
  Ning}, \bibinfo{person}{Zhen-Hui Liu}, \bibinfo{person}{Mu-Nan Ning},
  \bibinfo{person}{Yu-Yang Liu}, {and} \bibinfo{person}{Li Yuan}.}
  \bibinfo{year}{2023}\natexlab{}.
\newblock \showarticletitle{Llm lies: Hallucinations are not bugs, but features
  as adversarial examples}.
\newblock \bibinfo{journal}{{\em arXiv preprint arXiv:2310.01469\/}}
  (\bibinfo{year}{2023}).
\newblock


\bibitem[\protect\citeauthoryear{Zugecova, Macko, Srba, Moro, Kopal,
  Marcincinova, and Mesarcik}{Zugecova et~al\mbox{.}}{2024}]%
        {zugecova2024evaluation}
\bibfield{author}{\bibinfo{person}{Aneta Zugecova}, \bibinfo{person}{Dominik
  Macko}, \bibinfo{person}{Ivan Srba}, \bibinfo{person}{Robert Moro},
  \bibinfo{person}{Jakub Kopal}, \bibinfo{person}{Katarina Marcincinova}, {and}
  \bibinfo{person}{Matus Mesarcik}.} \bibinfo{year}{2024}\natexlab{}.
\newblock \showarticletitle{Evaluation of LLM Vulnerabilities to Being Misused
  for Personalized Disinformation Generation}.
\newblock \bibinfo{journal}{{\em arXiv preprint arXiv:2412.13666\/}}
  (\bibinfo{year}{2024}).
\newblock


\end{thebibliography}

\end{document}